
\input vanilla.sty

%

%

\def\\{{\tt\char'134}}
\def\{{{\tt\char'173}}
\def\}{{\tt\char'175}}
\def\@{\char'100}
%
%
%
\def\boxit#1{%
              \vbox{%
                    \hrule%
                    \hbox{%
                          \vrule%
                          \kern3pt
                          \vbox{\kern3pt\hbox{#1}\kern3pt}%
                          \kern3pt%
                          \vrule%
                          }%
                     \hrule%
                    }%
             }
\newdimen\hautbox \newdimen\largbox
\newbox\entbox
\def\shadowbox#1{
                 \setbox\entbox=\boxit{#1}
                 \hautbox=\ht\entbox
                 \largbox=\wd\entbox
                 \advance \hautbox by -2pt
                 \vbox{\hbox{\boxit{#1}\vrule width2pt height\hautbox}
                 \nointerlineskip\moveright 2pt \hbox{\vrule height2pt width
                 \largbox}\nointerlineskip}}
\def\border#1{%
              \vbox{%
                    \hbox{%
                          \kern3pt
                          \vbox{\kern3pt\hbox{#1}\kern3pt}%
                          \kern3pt%
                          \vrule%
                          }%
                     \hrule%
                    }%
             }
%

%
%
\def\fonthdg{c}

\def\aujourdhui{\space\number\day\space\ifcase\month\or
janvier\or f\'evrier\or mars\or avril\or
mai\or juin\or juillet\or ao\^ut\or septembre\or
octobre\or novembre\or d\'ecembre\fi
\space\number\year}

\def\today{\space\ifcase\month\or
January\or February\or March\or April\or
May\or June\or July\or August\or September\or
October\or November\or December\fi
\space\number\day,\space\number\year}
\def\it#1{{\sl #1\/}}

%
%
\newcount\nosection
\newcount\nosubsect
\newcount\nosubsub
\nosection=0
\def\section#1#2{\advance \nosection by 1 \nosubsect=0 \nosubsub=0
               \removelastskip\bigskip\leftline{\moyen #2{\number
               \nosection\ #1}}
               \medskip}
\def\subsection#1#2{\advance \nosubsect by 1 \nosubsub=0
               \removelastskip\medskip
               \leftline{\bf\hskip\parindent #2{\number\nosection.\number
               \nosubsect\ #1}}
               \medskip}
\def\subsubsection#1#2{\advance \nosubsub by 1
                  \removelastskip\medskip
                  {\bf\hskip\parindent #2{\number\nosection.\number\nosubsect.
                  \number\nosubsub\ #1}}
                  \medskip}
%
%
%

%
=\fonthdg mr8
=\fonthdg mtt10
=\fonthdg mr6
=\fonthdg mmi8 \skewchar\eighti='177
=\fonthdg mmi6 \skewchar\sixi='177
=\fonthdg msy8 \skewchar\eightsy='60
=\fonthdg msy6 \skewchar\sixsy='60
=\fonthdg mbx8
=\fonthdg mbx6
=\fonthdg msl8
\font\eightit=\fonthdg mti8
=\fonthdg mcsc10


\def\tenpoint{\Tenpoint}
\def\Tenpoint{\normalbaselineskip=12pt            
 \def\rm{\fam0\tenrm}%
 \def\it{\fam\itfam\tenit}%
 \def\sl{\fam\slfam\tensl}%
 \def\bf{\fam\bffam\tenbf}%
\def\smc{\tenrm}
 \def\mit{\fam 1}%
 \def\cal{\fam 2}%
 \textfont0=\tenrm   \scriptfont0=\eightrm   \scriptscriptfont0=\sixrm
 \textfont1=\teni    \scriptfont1=\eighti    \scriptscriptfont1=\sixi
 \textfont2=\tensy   \scriptfont2=\eightsy   \scriptscriptfont2=\sixsy
 \textfont3=\tenex   \scriptfont3=\tenex     \scriptscriptfont3=\tenex
 \textfont\itfam=\tenit
 \textfont\slfam=\tensl
 \textfont\bffam=\tenbf \scriptfont\bffam=\eightbf
   \scriptscriptfont\bffam=\sixbf
\setbox\strutbox=\hbox{\vrule height 8.5pt depth 3.5pt width 0pt}%
\def\tt{\tentt}\normalbaselines\rm}

\def\Eightpoint{\normalbaselineskip=10pt
 \def\rm{\fam0\eightrm}%
 \def\it{\fam\itfam\eightit}%
 \def\sl{\fam\slfam\eightsl}%
 \def\bf{\fam\bffam\eightbf}%
 \def\mit{\fam 1}%
 \def\cal{\fam 2}%
 \textfont0=\eightrm   \scriptfont0=\sixrm   \scriptscriptfont0=\sixrm
 \textfont1=\eighti    \scriptfont1=\sixi    \scriptscriptfont1=\sixi
 \textfont2=\eightsy   \scriptfont2=\sixsy   \scriptscriptfont2=\sixsy
 \textfont3=\tenex   \scriptfont3=\tenex     \scriptscriptfont3=\tenex
 \textfont\itfam=\tenit
 \textfont\slfam=\eightsl
 \textfont\bffam=\eightbf \scriptfont\bffam=\eightbf
   \scriptscriptfont\bffam=\sixbf
\setbox\strutbox=\hbox{\vrule height 7pt depth 3pt width 0pt}%
\normalbaselines\rm}
=\fonthdg mr10 scaled \magstep1
=\fonthdg mtt10 scaled \magstep1
=\fonthdg mmi10 scaled \magstep1 \skewchar\twli='177
=\fonthdg msy10 scaled \magstep1 \skewchar\twlsy='60
=\fonthdg mbx10 scaled \magstep1
=\fonthdg msl10 scaled \magstep1
\font\twlit=\fonthdg mti10 scaled \magstep1

\def\twlpoint{\Twlpoint}
\def\Twlpoint{\normalbaselineskip=14.4pt            
 \def\rm{\fam0\twlrm}%
 \def\it{\fam\itfam\twlit}%
 \def\sl{\fam\slfam\twlsl}%
 \def\bf{\fam\bffam\twlbf}%
 \def\mit{\fam 1}%
 \def\cal{\fam 2}%
 \textfont0=\twlrm   \scriptfont0=\tenrm   \scriptscriptfont0=\eightrm
 \textfont1=\twli    \scriptfont1=\teni    \scriptscriptfont1=\eighti
 \textfont2=\twlsy   \scriptfont2=\tensy   \scriptscriptfont2=\eightsy
 \textfont3=\tenex   \scriptfont3=\tenex     \scriptscriptfont3=\tenex
 \textfont\itfam=\twlit
 \textfont\slfam=\twlsl
 \textfont\bffam=\twlbf \scriptfont\bffam=\tenbf
   \scriptscriptfont\bffam=\eightbf
\setbox\strutbox=\hbox{\vrule height 8.5pt depth 3.5pt width 0pt}%
\def\tt{\twltt}\normalbaselines\rm}
%
%




%

\def\a{\alpha}

\def\l{\lambda}

\def\o{\omega}

\def\G{\Gamma}

\def\interligne{\baselineskip}
\def\cal{\Cal}

\def\ni{\noindent}

\def\abstract{\noindent{\bf Abstract : }}
\def\ack{\vskip 0.5cm\noindent\it Acknowledgments : \rm}

\def\sec#1{\advance \nosection by 1 \nosubsect=0 \nosubsub=0
               \removelastskip\bigskip\leftline{\bf\number\nosection.\ #1}
               \medskip}
\def\subsec#1{\advance \nosubsect by 1 \nosubsub=0
               \removelastskip\medskip
               \leftline{\bf\hskip\parindent{\number\nosection.\number
               \nosubsect\ #1}}
               \medskip}
\def\subsubsec#1{\advance \nosubsub by 1
                  \removelastskip\medskip
                  {\bf\hskip\parindent{\number\nosection.\number\nosubsect.
                  \number\nosubsub\ #1}}
                  \medskip}

\def\REF{\newpage{\bf\centerline{REFERENCES}}\vskip 2cm}
\def\FIG{\newpage{\bf\centerline{FIGURE CAPTIONS}}\vskip 2cm}
\newcount\noref
\newcount\nofig
\newcount\noeq
\noeq=0
\nofig=0
\noref=0
\def\ad{\advance \noeq by 1}

\def\ref{\advance \noref by 1\item{[\number\noref]}}
\def\fig{\advance \nofig by 1\item{Fig. \number\nofig\ :}}
\hsize 16.2cm \vsize 23cm
\twlpoint
\TagsOnRight
%
%

\def\xx{\bold X}
\def\G{\Gamma}
{\bf\centerline{DECAYING TURBULENCE AND THE DYNAMICS OF}
{\bf\centerline{DIFFUSING VORTICES WITH CONSERVATION LAWS}}

\interligne 0.8cm
\vskip 2.1cm
\smc\centerline{Cl\'ement Sire\rm
\footnote{\tenpoint E-mail: clement\@siberia.ups-tlse.fr.\twlpoint}}

\vskip 1cm
\it\centerline{Laboratoire de Physique Quantique
(Unit\'e associ\'ee 505 du CNRS)}
\centerline{Universit\'e Paul Sabatier, 31062 Toulouse Cedex, France}
\rm

\interligne 24pt
\vskip 2cm
\ni {\bf Abstract:} In this letter, I solve a model for the
dynamics of vortices in a decaying two-dimensional turbulent fluid.
The model describes their effective diffusion, and the merging
of pairs of vortices of same vorticity sign, when they get too close.
The merging process is characterized by the conservation
of energy and of the quantity $Nr^n$, where $r$ is the mean vortex radius,
and $N$ their number. $n=4$ corresponds to a constant peak vorticity,
and $n=2$ to a constant kurtosis.
I found the scaling laws for various physical quantities ($r$,
enstrophy, kurtosis...), and for instance, it is shown that
$N\sim (t/\ln(t))^{-\frac{2n}{3n-4}}$ for $n>2$, and $N\sim t^{-2}$
for $n=2$, in good agreement with extensive numerical
simulations.
I also discuss some recent experiments in view of these results.

\vskip 1.2cm
\ni PACS numbers: 47.27.-i, 82.20.-w, 05.40.+j.

\ni Short title: Dynamics of vortices.


\newpage


Recent experimental [1-2] and theoretical [3-7] works have emphasized
the importance of coherent vortex dynamics during the decay
of turbulence in $d=2$.
This process consists mainly in three stages: during an initial transient
period,
the fluid self-organized and a network of coherent vortices appears. Once the
coherent vortices have emerged, vortices disappear essentially through merging
of same sign vortices, such that their number decreases and their average size
increases, in a process somewhat reminiscent of a coarsening stage
[8][9].
Finally, when only one (or very few)
dipole is left, it decays diffusively, due to the finite viscosity.

In order to describe this second stage of the dynamics of decaying turbulence,
the authors of [6][7] introduced the following simple model.
The starting point lies in the well known fact that
the evolution dynamics of a set of far apart vortices is a conservative
process, such that vortex centers positions $\{\xx_i\}$ evolve according
to Kirchoff laws [10],
$$
\G_i\frac{d\xx_i}{dt}=\nabla_i\times\Cal H, \qquad
\Cal H=-\sum_{i,j}\G_i\G_j\ln|\xx_i-\xx_j|\tag 1
$$
where the circulation $\G_i=\pi r_i^2\o_i$ is the area of the $i$-th vortex
multiplied by its average core vorticity.
In addition to the dynamics, one must describe the dissipation process
when some vortices get too close. Experiments [1][2] or numerical simulations
[5-7] have consistently shown that this happens mainly
through a complicated merging process between vortices of $same$ vorticity
sign,
while dipoles form a very stable state [2]. In [6-7], this was modelized by
assuming that when two vortices with radii $r_1$ and $r_2$ are at a distance
less than a $d_c(r_1,r_2)$, they instantly merge.
The authors of [6] found that a consistent form for this minimal
approach distance is $d_c(r_1,r_2)\approx 2.592r_2+0.609r_1^2/r_2$, for
$r_1\le r_2$. In the present paper, the claim is made that the precise form of
$d_c$ is not important as far as the critical properties of the model are
concerned, provided $d_c$ is of order of the mean vortex radius $r$.
When two vortices of same sign merge, we still need to
define the properties of the resulting
vortex. Motivated by their numerical results for the full Navier-Stokes
equation, the authors of [5-6] were lead to impose local conservation of energy
and a time independent typical peak
vorticity $\o$. This was also found to be consistent with the experiment
by Tabeling et al. [1]. Since the energy of a vortex
is of order $\G^2\sim \o^2 r^4$, the new vortex has
a radius $r_3$ satisfying, $r_3^4=r_1^4+r_2^4$.
On the other hand, following the observation of the experiment described in
[2] (see below), we could alternatively impose that the energy and
the total area occupied by vortices remains constant, which
is essentially equivalent to have a constant kurtosis $K$.
The new merging rule is then, $r_3^2=r_1^2+r_2^2$, and
$\o$ must decay in order to keep the energy constant.
More generally, we will treat the general case of a dynamics with the
conservation of the local ``charge'' $q_i=r_i^n$.

We now derive the scaling relations between the number of vortices $N$,
their mean radius $r$, their typical distance $l$,
the ratio $Z/E$ ($Z$ is the enstrophy), the kurtosis $K$, and
$\o/\sqrt{E}$. For large time,
we expect that $r\sim t^\nu$ [1-2][5-7][11][12].
The assumed conservation law, and the fact
that $E\sim N\o^2 r^4$, $Z\sim N\o^2r^2$, $K\sim N r^2$, leads to,
$$
\aligned
r\sim t^\nu,\quad l\sim & t^{\nu\frac n 2},\quad N\sim t^{-n\nu},\quad \\
\frac{Z}{E}\sim t^{-2\nu},\quad K\sim &t^{(2-n)\nu},
\quad \frac{\o}{\sqrt E}\sim t^{-(2-\frac n 2)\nu}
\endaligned\tag 2
$$
which was obtained in [5-6] for the case $n=4$. We need $n\geq 2$ to keep
$l\geq r$, while for $n<2$, one can show that $N\sim N_0\exp(-t/t_0)$ [9].
We see that the problem now amounts to determine the unknown
exponent $\nu$ which should depend explicitly on the conservation
law considered in the dynamics.
Ultimately, one should be able to infer on physical grounds
(role of initial conditions [7], viscosity, or $d=3$ effects [2]) which
effective conservation law, or $n$, is relevant in experiments or numerical
simulations of (quasi-) two-dimensional decaying turbulence.
I now focus on the relation between $\nu$ and $n$,
and sketch a rather simple and physical derivation for it, while
a more elaborate solution is possible (see conclusion).
This solution is based on the
mapping of the original problem on a solvable dynamics for a
self-organized system [13-14]. This, maybe, gives a more
solid justification for the description of the coherent vortex structure in
terms of a self-organized system [15], as mentioned by some authors [6].

The first step of the solution uses the claim that the motion of
vortices in the absence of merging is effectively diffusive,
which should result from the chaotic nature of their motion.
{}From Kirchoff equations, and using the fact that the total
circulation is zero, the mean square velocity
of vortex centers is,
$v^2=\left\langle\sum_j\frac{\G_j^2}{R_{ij}^2}\right\rangle\sim
N\o^2r^4\int_r^L\frac{R\,dR}{R^2}\sim E\ln(L/r)$,
where the upper cut-off is the linear size of
the system $L$, and the lower one is the mean vortex radius
$r$. Note that $v^2$ is strictly constant with the correct definition of
energy $E\sim N\o^2r^4\ln(L/r)$, but decays slowly with the definition
$E\sim N\o^2r^4$ considered in the simulations of [6] and also here, for the
moment. The diffusion constant of a vortex is then of order $D\sim v^2/\o$,
which $varies$ in time if $\o$ is not constant.
We can now consider separately the two gas of vortices with positive and
negative vorticity, since the effect of one on the other is taken into
account in the diffusion constant, and since mergings only involve
same sign vortices [2][5-7]. It now remains to treat
correctly the apparently complicated merging condition. This
can be done by mapping the
problem to an effective dynamics on a lattice of suitable constant $a$,
and by choosing consistently a time step $\tau=a^2/D$. If one takes $a=r$,
the merging process is correctly described by saying that two charges add
together when they hop on the same lattice site. Indeed, in this case
their mutual distance is then less than the mean vortex
radius, which is in accordance with the kind of merging conditions
introduced above. We have thus separated the dynamics in two processes: $(i)$
$free$ diffusion of far apart charges (vortices), with a diffusion constant
$D\sim v^2/\o$, and $(ii)$ addition (merging)
of charges hopping on the same site (vortices at a distance less that
$a=r$). What makes the problem solvable is the fact that the dynamical
equations for the time evolution of the $q_i$'s are now linear:
$$
q_i(t+\tau)=\sum_{|i-j|=a}w_{j\to i} q_j(t), \qquad w_{j\to i}=0\text{ or }1,
\quad\sum_{i}w_{j\to i}=1
\tag 3
$$
where the $w_{j\to i}$'s are random variables, and
the last condition expresses that the vortex and its associated charge
on site $j$ hops on one of its neighboring sites (with equal probabilities).
This problem can be solved exactly [13-14], even in the case when one
adds a random charge $I_i(t)$ to the left hand side of Eq. (3),
which could simulate the driving noise
in fully developed turbulence [9]. For the purpose of this letter,
we do not need the full solution of Eq. (3),  but only need to know that the
upper critical dimension beyond which mean-field theory is exact is
$d=2$ [16][14]. Assuming no
correlation (mean-field), the number of collisions at time $t$
is proportional to $N^2$. Then, $\frac{dN}{dt}\sim -(a/L)^d N^2/\tau$, which
immediately leads to $N\sim (L/a)^d(t/\tau)^{-1}$, where $L$ is the linear
size of the system and $d>2$ is the space dimension.
For $d=2$ and for $n>2$ ($l/r$ then diverges when $t\to\infty$ and vortices are
point-like objects), one can show exactly [17] the existence of logarithmic
corrections in this kind of problems, such that
$N(t)\sim L^2\ln(t/\tau)/(Dt)$.

In order to illustrate the quantitative effect of such a
correction on the numerical determination of the decay exponent,
I show in fig. 1 numerical simulations of Eq. (3) with a constant
$D=a^2/\tau$ on a $1500\times 1500$ square lattice with all sites
initially occupied, and up to $t=10^5\tau$.
The best fit by the law $N(t)=N_0(1+t/t_0)^\a$ (as used
in [6]) gives a remarkably linear fit with $\a= 0.90\pm 0.01$ even after
five decades in time (see fig. 1-a)! Of course, the plot of $tN(t)$ $vs$
$\ln(t)$ immediately reveals the logarithmic correction (fig. 1-b).
This remark motivates my next comment:
there have been some claims that the vortex dynamics in
decaying turbulence presents some similarities with the coarsening dynamics
of the $XY$ model [18], for which the number of spin vortices was
apparently seen numerically to decay as $N\sim t^{-0.75}$ [18-19],
as in the Kirchoff vortex model [6]. Recent more extensive simulations [20],
have shown that this apparent behavior is
due to an important logarithmic correction, and that $N\sim\ln(t)/t$,
as expected from the theory of coarsening systems with non conserved
order parameter [8][20]. Since the anomalous exponent $\xi$ seen in the
$XY$ model seems to be an artifact of unanticipated
logarithmic corrections, it is natural to challenge the observation
of such exponents in the field of decaying turbulence. Notice however that
the vortices in the $XY$ model behave in a qualitatively quite different
fashion as in a
turbulent fluid [9]. First of all, they are quantized, and after a short
transient time, only vortices with winding number $\pm 1$ remain [20][8].
Contrary to the merging dynamics of fluid vortices already described,
that of the $XY$ model only involves the $annihilation$ of $opposite$
(quantized) vortices.
After these preliminary remarks, I end the derivation of the scaling
exponent $\nu$, and confirm the role of logarithmic corrections.
Using the scaling equations Eq. (2), and the above result for $N(t)$, one
easily finds, $r\sim (v^2t/\ln(t))^{\nu}$, with
$\nu=\frac{2}{(3n-4)}$, since $v$ does not behaves faster than a logarithm
in time.

For the first case of a dynamics with conserved energy and constant $\o$
($n=4$) studied in [5-6][7],
the result is $r\sim(v^2 t/\ln t)^{1/4}$, with $v=\ln(t_{end}/t)$, and
where $t_{end}$ is of the order of the time for which only one
(pair of positive and negative) vortex is left in the sample. We thus find
$\nu=1/4$, or in the language of [5-6], $\xi=4\nu=1$.
The authors in [6] found $\xi= 0.72\pm 0.02$. I will now show that the
discrepancy with the present analytical result is certainly
due to the two logarithmic corrections, the first one in $v$, and the second
one
coming from the critical $d=2$ corrections, which both tend to $lower$ the
effective exponent $\xi$. In fact, it would have been more correct to
consider the definition $E\sim N\o^2r^4\ln(L/r)$ [6], taking
into account the long
range interactions between vortices, leading to the local conservation of
$r^4\ln(L/r)$. This implies that the typical vortex square velocity $v^2$
is now $constant$. Still, there remains the most important logarithmic
correction
(due to $d=2$), which in principle, should be present in the numerical
solution of the full Navier-Stokes equation, for which the authors of [6]
also found $\xi\approx 0.70\sim 0.75$.
In fig. 1, we show the result of numerical simulations of the
problem of diffusing vortices merging when their distance is less
than $d_c(r_1,r_2)=r_1+r_2$, and with local conservation of $r^4$,
but with a constant $v^2$ (and $D$ since $\o$ is constant for $n=4$),
in order to avoid the introduction of extra
logarithmic corrections (present in [6]). One finds $\xi=4\nu=0.84\pm 0.01$
(fig. 1-a), from a fit to the form $N(t)\sim N_0(1+t/t_0)^{-\xi}$ as used
in [5-6], even when using much bigger samples than in [6-7] ($N_0= 20000$
instead of $N_0\sim 400$ in [6] and $N_0=300$ in [7]), which is allowed by the
simpler nature of the model. However, for this model, the exact solution
presented here predicts $N(t)\sim \ln(t)/t$, which is illustrateded on
fig. 1-b by plotting $tN(t)$ $vs$ $\ln(t)$.
I now present numerical simulations of the full model introduced in
[6-7], where the vortices evolve according to Kirchoff dynamics (Eq. (1)),
and for which the same merging criterion as in [6] as been used.
The calculations are much heavier than for the simple diffusing
vortex model, which drastically limits the number of vortices and integration
time that one can simulate. In order to obtain the properties of the scaling
regime, the authors of [6] introduced a clever
renormalization scheme: they have simulated the decay of the system
from 400 to 100 vortices, and used the final configuration to
generate a new initial condition with 400 vortices, by
copying it four times. They repeated this renormalization scheme until they
obtained an apparent scaling regime and enough statistics. The drawback
of this method is that it only allows the precise measurement of the decay
of the system during less than a decade in time, during which the number
of vortices is divided by a factor 4 (in fact less [6][9]; see also fig. 1).
They did not find a pure power law decay, but found a good fit to
$N(t)\sim N_0(1+t/t_0)^{-\xi}$, with $\xi\approx 0.72$. This form
simulates the effective (sharp) increase with time of the exponent $\xi$,
as they found quite large values for $t_0\sim t_{end}/3$
($t_{end}$ is the final time of the simulation).
This scaling was shown in [11] to be consistent with the
assumption that the population of vortices is at all time in local Boltzmann
equilibrium. Note however that this assumption would be wrong in a coarsening
dynamics (for instance for the $XY$ model), for which the correlation functions
during the dynamics are $unrelated$ to the equilibrium ones [8][9][20].
Such a scaling with roughly the same effective $n=4$ and $\xi\approx 0.7$
was also observed experimentally in [1]. As far as the
direct numerical simulation of the Navier-Stokes equation is concerned, the
authors of [6] claim to see the same behavior as described above,
while some other groups discuss the existence of such growth exponents
[21], or their values [22].
The simulations presented in fig. 1 use the same
method as in [6], but with an initial and final number of vortices
respectively equal to 1800 and 200 (instead of 400 and 100).
The details will be given in [9]. The apparent
exponent as extracted by the method of [6] is $\xi=0.76\pm 0.02$ (fig. 1-a),
systematically larger than in [6]. In [7], where smaller samples
($N=300$) and shorter times than in [6] where considered, the authors found
an even lower $\xi\sim 0.6$.
In fig. 1-b, I show the plot of $tN(t)$ $vs$ $\ln(t)$, which is found to be
reasonably linear.
Note that logarithmic
corrections as found here due to the merging (coarsening) dynamics
of point-like objects ($r\ll l$) in $d=2$ could also be generated
by taking into account [9]
the long range interaction between vortices as in the $XY$ model [20].
Finally, the mean-field vortex radii distribution can be computed [9]
using Eq. (3) and the results of [14]. Once radii are expressed
in unit of the average radius $r$, the result is
$P(x)=4\l x^3\exp(-\l x^4)$, where $\lambda=\G(5/4)$. The distribution
is shown on fig. 2.
The agreement with numerical simulations is satisfactory, the slight
discrepancy being probably due to the difference between the merging distance
used in the simulations ($d_c=r_1+r_2$) and in the exact calculation ($d_c=r$),
and the fact that some small correlation (non mean-field) effects should
be present in $d=2$. The distribution for the Kirchoff dynamics
is slightly wider [6], but $d_c$ is probably too different from the
merging condition in the solvable model.

We now study the case of the coherent vortex dynamics
with conserved kurtosis.
In other words, the area covered by the vortex cores remains constant.
This phenomenologically reproduces the observed
feature of the experiment presented in [2].
If we also impose conservation of energy, the maximal vorticity must
then decay as $\o\sim r^{-1}$ according to the relation $E\sim
Nr^4\o^2\sim r^2\o^2$, for constant $K$. Noting that $D\sim v^2/\o\sim r\sim l$
increases with time, one finds
$r\sim v^2t$, such that $\nu=1$, with no logarithmic corrections (for constant
$v$) since $r\sim l$, and thus vortices
are no more point-like objects [9].
This exactly reproduces Batchelor
phenomenology [12], in the framework of a $microscopic$ theory. I have
simulated Kirchoff's dynamics for $n=2$, with initially $N=1600$ vortices.
After one renormalization step, $N$ is left decaying. In this case the dynamics
is much faster than for $n=4$, which permits to observe the decay up to
$N\sim O(1)$ (fig. 3). The fit to the form $N(t)=N_0(1+t/t_0)^{-2}$ is
excellent (fig. 3-a). Similar results are obtained (on more decades)
for the  diffusing vortex model with $D\sim r$, confirming $\nu=1$ (fig. 3).
In the experiment described in [2], the obtained
scaling relations are exactly of the form given in Eq. (2) with $n=2$,
but the experimental estimate for $\nu$ (in the first time decade)
lies in the range  $\nu\approx 0.18\sim 0.26$. In fact, a naive
determination of $\nu$ in a Kirchoff vortex
simulation with initially $N_0=100$ vortices,
in the time range for which $0.2N_0\leq N\leq 0.8N_0$ (as in
[2]), leads to $2\nu=0.54\pm 0.04$ in
surprisingly good agreement with experiment (see fig. 3 with $N_0=1600$).
Note that three-dimensional effects, leading
to the diffusion of vorticity in the third dimension, could be
at the origin of the constant area occupied by vortices in this
experiment [2].

In conclusion, I have solved a simple model of diffusing vortices,
with the effective diffusion constant derived from the Kirchoff dynamics
and conservation laws. I have found long transient regimes (due to
logarithmic corrections for $n>2$)
which, I argue, are at the origin of the smaller exponents observed in the
literature. For a constant peak vorticity $\o$ ($n=4$),  vortices
are effectively diffusive ($D\sim v^2/\o$ is then constant), while for a
constant kurtosis $K$ ($n=2$), their motion is effectively
ballistic ($D\sim t$). In the presence of a finite viscosity (diffusive
dissipation),
one could expect that the first dynamics is more stable than the Batchelor
regime, the motion of ballistic vortices being more easily damped than
that of diffusing ones.
Note finally that the problem of diffusing vortices can be solved in many
other situations: finite driving noise, initial long range distribution
of vortex radii (as numerically studied in [7]),...
This, and more details on the
numerics, as well as a new field theory for $d=2$ decaying turbulence in the
spirit of the vortex model will be presented in [9].

\ack I am grateful to Y. Pomeau and P. Tabeling, for interesting discussions,
and S.N. Majumdar for sharing so many interests with me. D.A. Huse is
acknowledged for precisions concerning [18-20].

\newpage

\REF
\item{[1]} P. Tabeling, S. Burkhart, O. Cardoso, H. Willaime, Phys. Rev. Lett.
{\bf 67}, 3772 (1991). A recent experiment by D. Marteau, P. Tabeling,
O. Cardoso (Preprint ENS/94), also suggests a conservation of peak vorticity.
\item{[2]} O. Cardoso, D. Marteau, P. Tabeling, Phys. Rev. E {\bf 49}, 454
(1994).
\item{[3]} J.C. McWilliams, J. Fluid. Mech. {\bf 146}, 21 (1984).
\item{[4]} M.E. Brachet, M. Meneguzzi, H. Politano, P.L. Sulem,  J. Fluid.
Mech. {\bf 194}, 333 (1988).
\item{[5]} G.F. Carnevale, J.C. McWilliams, Y. Pomeau, J.B. Weiss, W.R. Young,
Phys. Rev. Lett. {\bf 66}, 2735 (1991).
\item{[6]} J.B. Weiss, J.C. McWilliams, Phys. Fluids A {\bf 5}, 608 (1993).
\item{[7]} R. Benzi, M. Colella, M. Briscoloni, P. Santanglo, Phys. Fluids A
{\bf 4}, 1036 (1992).
\item{[8]}  Gunton J.D., San Miguel M., Sahni
P.S., in \it Phase Transitions and Critical Phenomena\rm, Eds. C. Domb and
J.L. Lebowitz (Academic, NY 1989), Vol. 8, p. 269; A.J. Bray, NATO ASI
on \it Phase Transitions and Relaxation in Systems with Competing Energy
Scales\rm, Geilo, Norway (1993); A.J. Bray,  Advances in Physics
(1995).
\item{[9]} C. Sire, in preparation; the methods developed in
C. Sire, S.N. Majumdar,  Phys. Rev. Lett. (1995, in press),
Phys. Rev. E (1995, in press), are applied to
the problem of $d=2$ decaying turbulence.
\item{[10]} G. Kirchoff, in \it Lectures in Mathematical Physics,
Mechanics \rm (Teubner, Leipzig, 1877).
\item{[11]} Y. Pomeau,  in \it Turbulence: A Tentative Dictionary\rm, Eds.
P. Tabeling and O. Cardoso (Plenum Press, NY 1995); Communication at the
conference \it Turbulencia\rm, Ed. M.G. Velarde (Aguadulce, Spain 1993).
\item{[12]} G.K. Batchelor, Phys. Fluids Suppl. II {\bf 12}, 233 (1969).
\item{[13]} H. Takayasu, M. Takayasu, A. Provata, G. Huber, J. Stat. Phys. {\bf
65}, 725 (1991).
\item{[14]} S.N. Majumdar, C. Sire, Phys. Rev. Lett. {\bf 71}, 3729 (1993).
\item{[15]} P. Bak, C. Tang, K. Wiesenfeld, Phys. Rev. Lett. {\bf 59},
381 (1987); D. Dhar, Phys. Rev. Lett. {\bf 641}, 1613 (1990).
\item{[16]} D. Dhar, private communication (1993).
\item{[17]} $N(t)$ decays as the inverse of the number of distinct
sites visited after time $t$ [9]; see e.g. C. Itzykson, J.-M. Drouffe,
\it Th\'eorie Statistique des Champs \rm (Inter Editions/Eds. CNRS, Paris
1989);
for related problems, see F. Leyvarz, S. Redner, Phys. Rev. A {\bf 46}, 3132
(1992).
\item{[18]} G. Huber, P. Alstrom, Physica A {\bf 195}, 448 (1993).
\item{[19]} M. Mondello, N. Goldenfeld, Phys. Rev. A {\bf 42}, 5865 (1990).
\item{[20]} B. Yurke, A.N. Pargellis, T. Kovacs, D.A. Huse,
Phys. Rev. E {\bf 47}, 1525 (1993).
\item{[21]} W. H. Matthaeus, W.T. Stirbling, D. Martinez, S. Oughon, D.
Montgomery, Phys. Rev. Lett. {\bf 66}, 2731 (1991); Physica D {\bf 51}, 531
(1991).
\item{[22]} D.G. Dritschel, Phys. Fluids A {\bf 5}, 984 (1993).

\newpage

\FIG

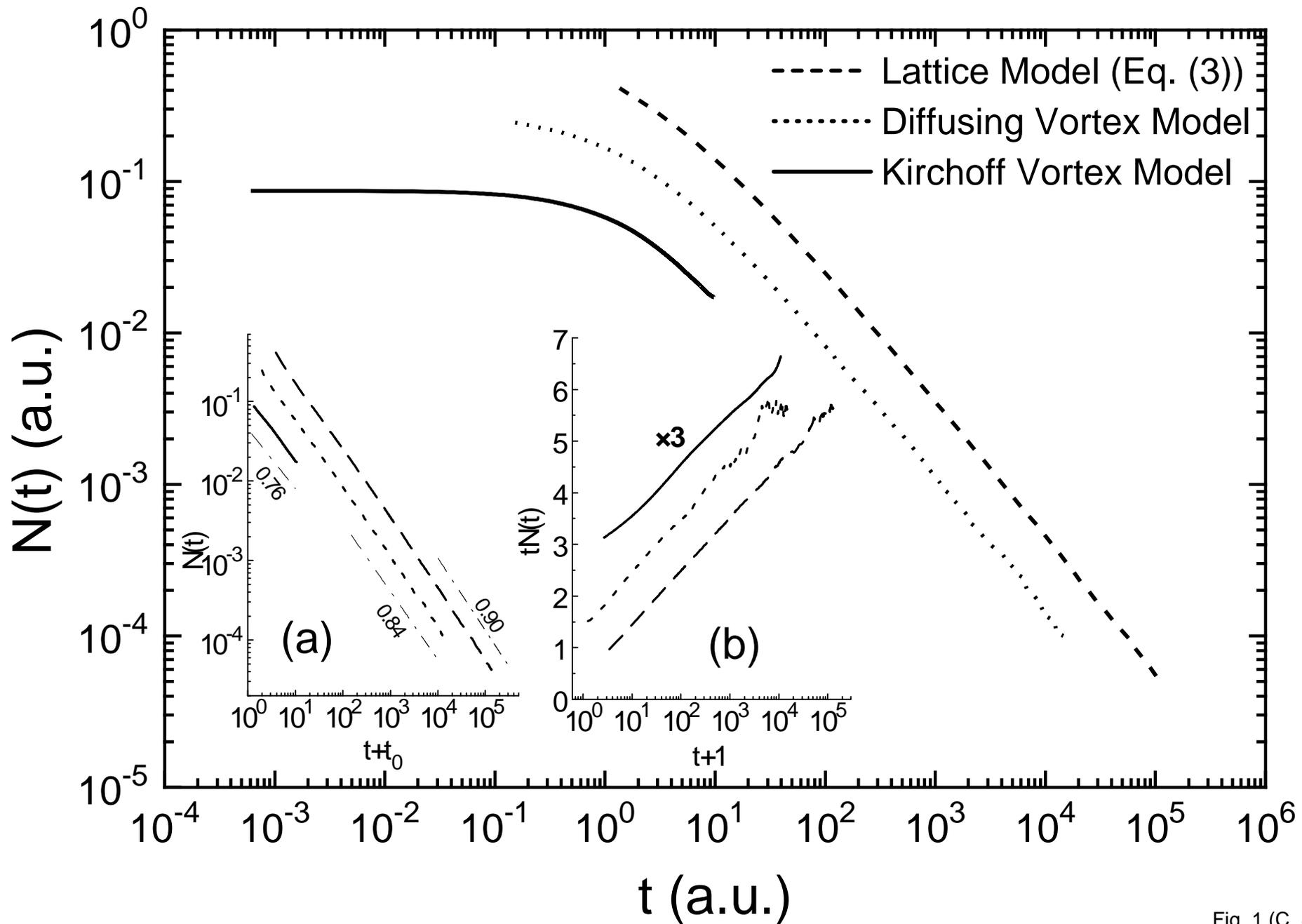
\item{Fig. 1:} Number of remaining vortices $N(t)$ after a time $t$ for the
different models introduced in the text ($n=4$). The time and number scales are
arbitrary and have been chosen appropriately in order to compare the
effective simulation time ranges. The initial numbers of particles are
respectively $N=2.25\times 10^6$, $N=20000$ and $N\sim 1500<1800$ due to
mergings right after a renormalization step [6][9]
(respectively 4, 10 and 10 samples).
(a) fits to $N(t)=N_0(1+t/t_0)^{-\xi}$ and the
effective values for $\xi$. (b) $tN(t)$ is now plotted $vs$ $\log(1+t)$.
For clarity,
the curve for the Kirchoff dynamics has been enlarged by a factor 3, due
to the much smaller time range, and the different curves have been shifted.

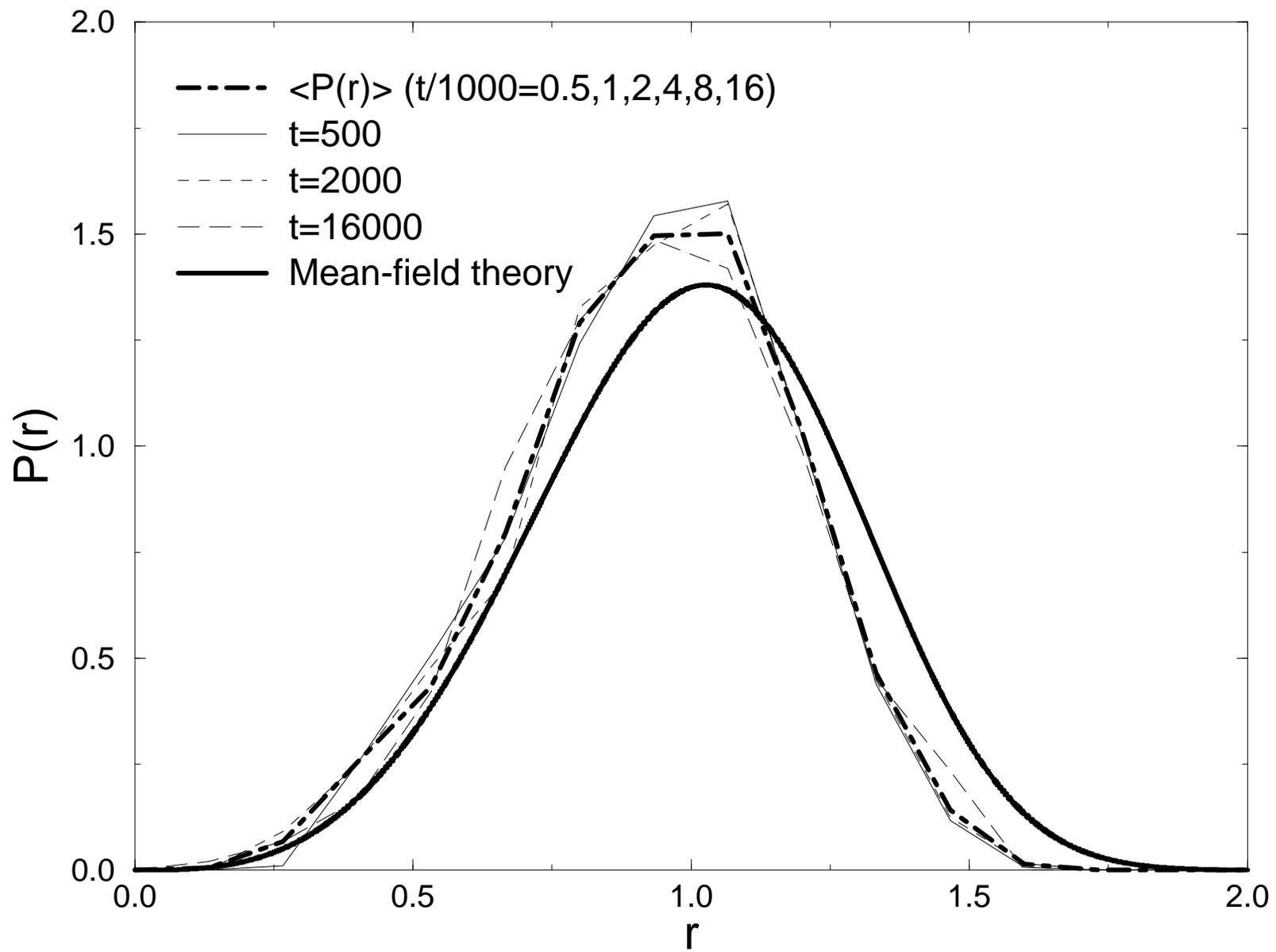
\item{Fig. 2:} Normalized vortex radii distributions obtained from the
diffusing
vortex simulation for different times. This is compared to the exact
mean-field one (see text).

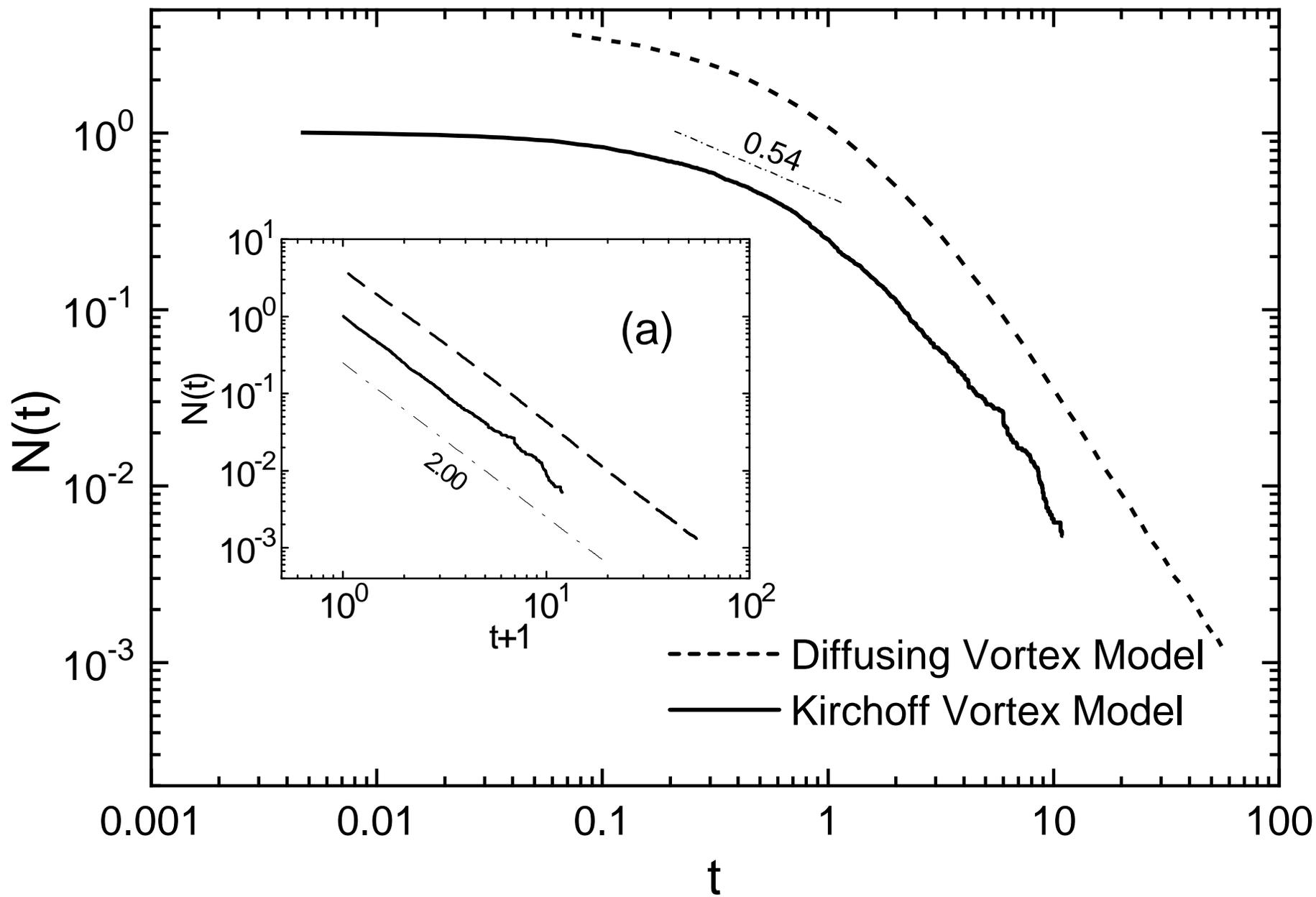
\item{Fig. 3:} For the Batchelor case ($n=2$), $N(t)$ is shown for
the diffusing vortex model (40 samples) and Kirchoff dynamics (10 samples).
(a) In the same units, $N(t)$ is plotted $vs$ $(1+t)$.

\end